\documentclass[11pt]{article}
\usepackage{moriond,epsfig}

\def\be{\begin{equation}}
\def\ee{\end{equation}}
\def\bea{\begin{eqnarray}}
\def\eea{\end{eqnarray}}

\begin{document}
\vspace*{4cm}
\title{$B \to$ Baryon decays in Belle}

\author{ S.-W. Lin }

\address{Physics Department, National Taiwan University, Taipei 106, Taiwan}

\maketitle\abstracts{
We report recent observations of baryonic $B$ decays with charmless and
charmed baryons in the final state. We show the angular distributions of the di-baryon low-mass enhancements in the charmless three-body baryonic $B$ decays and
the branching
fractions of $B$ decays with two charmed baryons or charmonium 
in the final states. We also report the observation of the decay $\eta_c \to \Lambda \bar{\Lambda}$ at Belle.
}
\section{Introduction}
\begin{figure}
\centering
\epsfig{figure=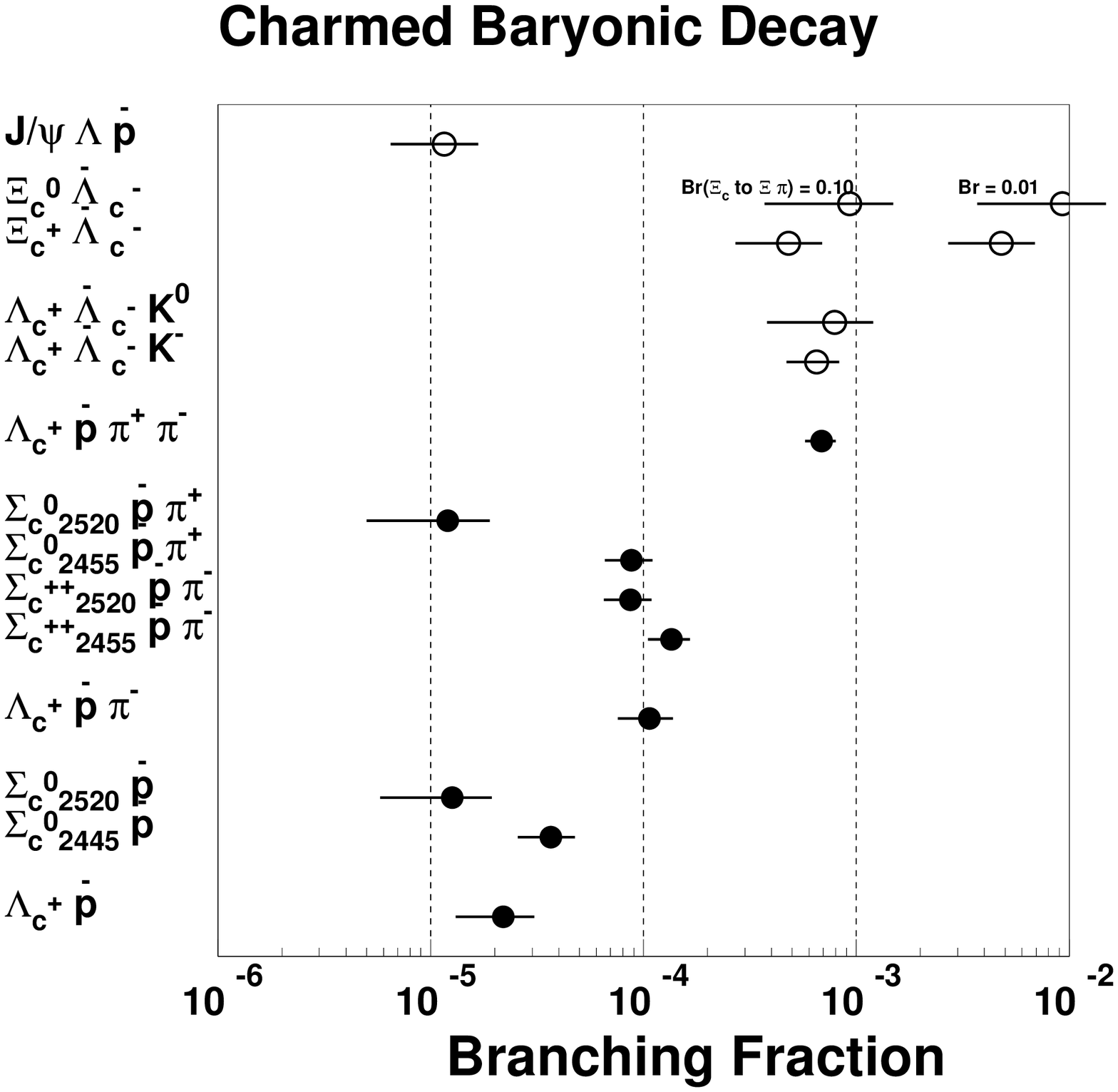,height=1.5in}
\epsfig{figure=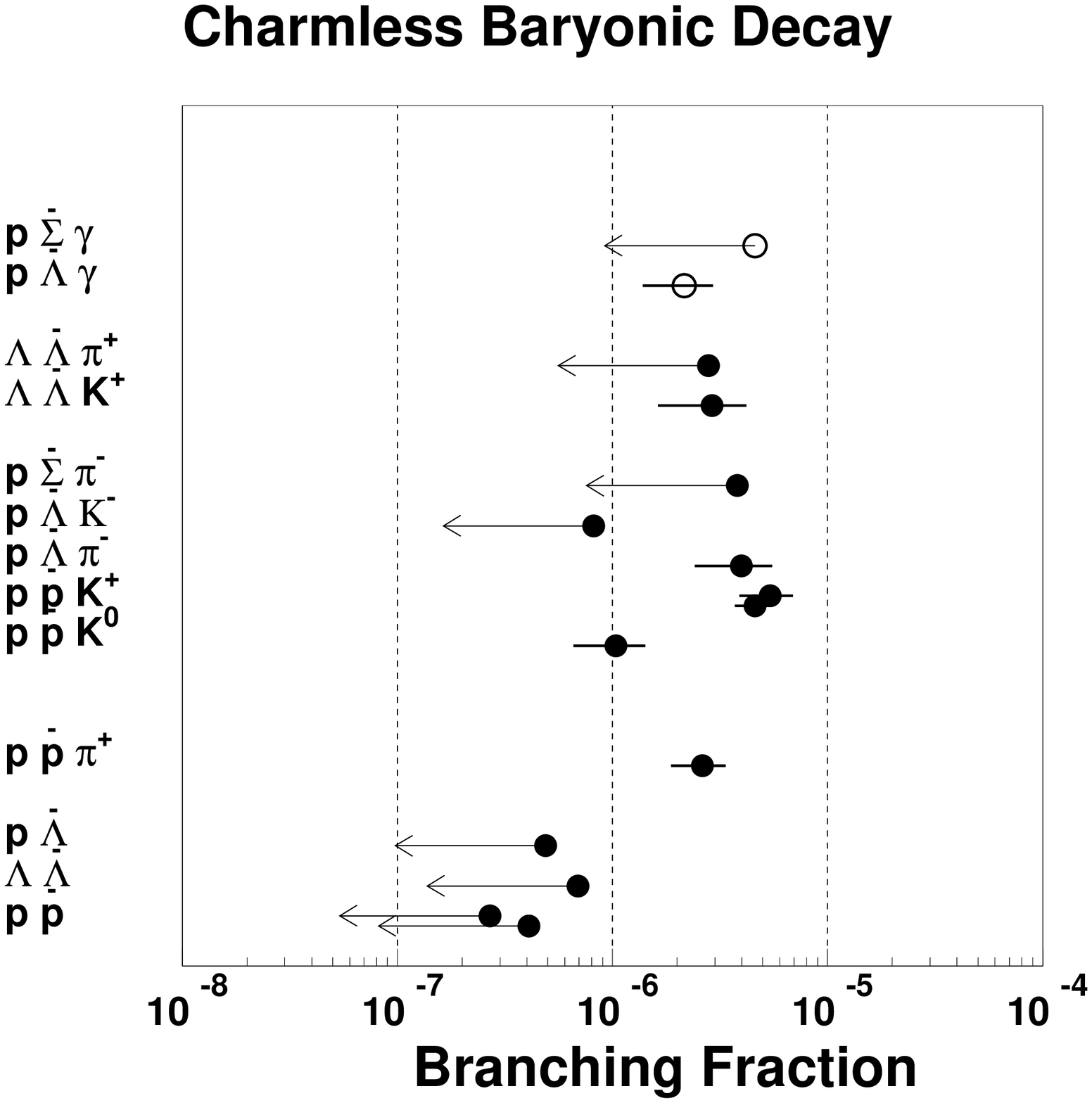,height=1.5in}
\caption{Branching fractions for charmless and charmed baryonic $B$ decays. The
open circles indicate new measurements  from Belle.
\label{fig:br}}
\end{figure}

Observations of several baryonic $B$ decays have been reported by Belle. The
measured branching fractions for charmless and charmed baryonic $B$ decays are 
shown in Fig. \ref{fig:br}.
In the charmless final states, only the three-body decays have been observed\cite{bar1,bar2,bar3,bar4,bar5}. In this contribution, we report on the angular distribution of the di-baryon low-mass 
enhancements seen in the charmless three-body baryonic $B$ decays. The data support the quark
fragmentation interpretation, while the gluonic resonance picture 
is disfavored. In the charmed final states, we observed the two-body and 
three-body decays with two charmed baryons or charmonium.
From the latter we can extract the branching fractions of $\eta_c$ into baryon pairs $p \bar{p}$ and - for the first time - into $\Lambda \bar{\Lambda}$.
Measuring decay rates of $\eta_c$ to different di-baryon modes is
a very useful check for theoretical predictions \cite{ans} and can shed light on
quark-diquark dynamics. The data sample was collected with Belle detector at the
KEKB asymmetric-energy $e^+ e^-$(3.5 GeV on 8.0 GeV) collider\cite{kekb}. KEKB 
operates at the $\Upsilon(4S)$ resonance ($\sqrt{s}$ = 10.58 GeV) with a peak 
luminosity that has exceeded 1.5 $\times 10^{34}$ cm$^{-2}$s$^{-1}$.

\section{Charmless baryonic $B$ decays}
\begin{figure}
\centering
\epsfig{figure=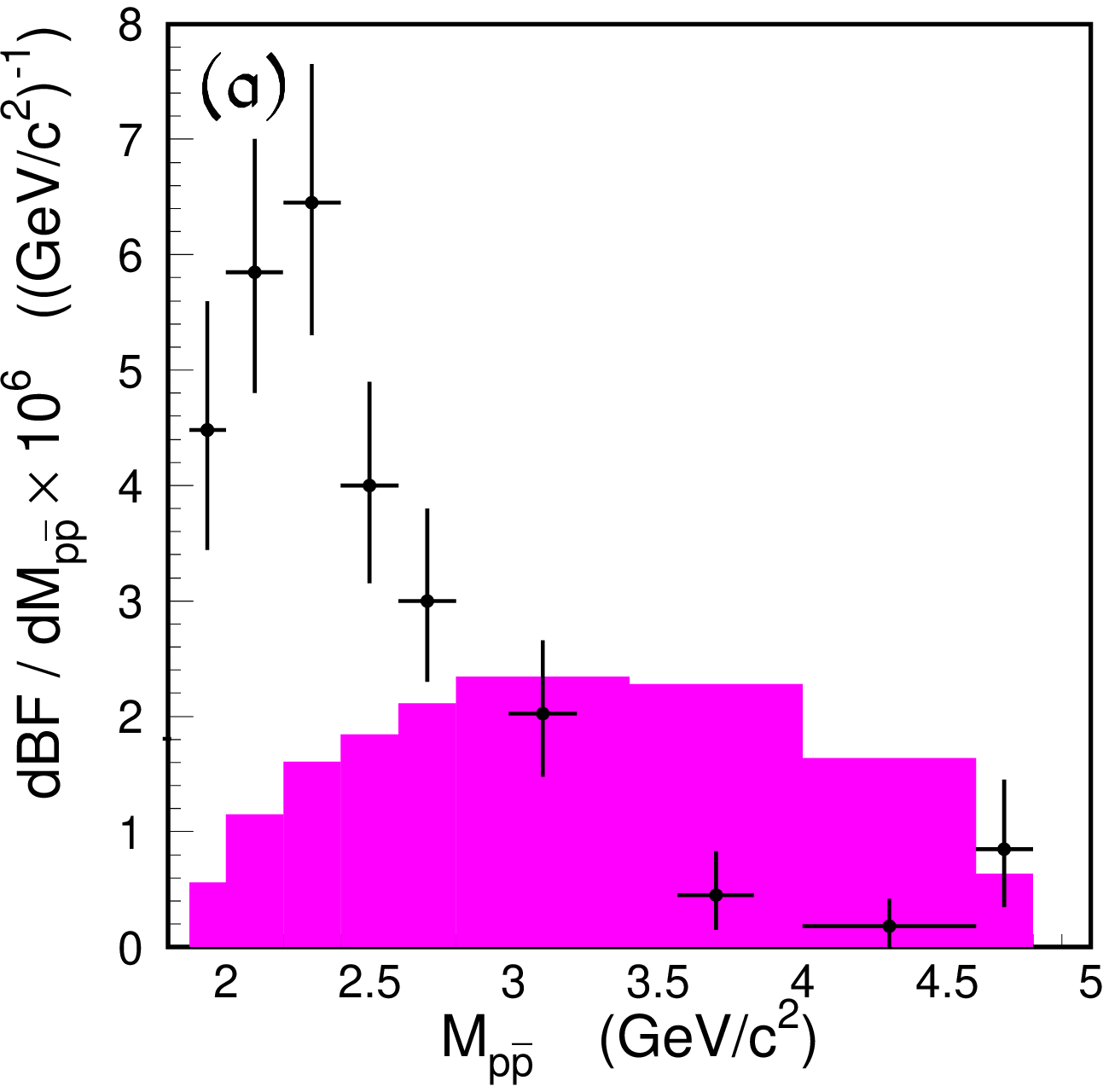,height=1.5in}
\epsfig{figure=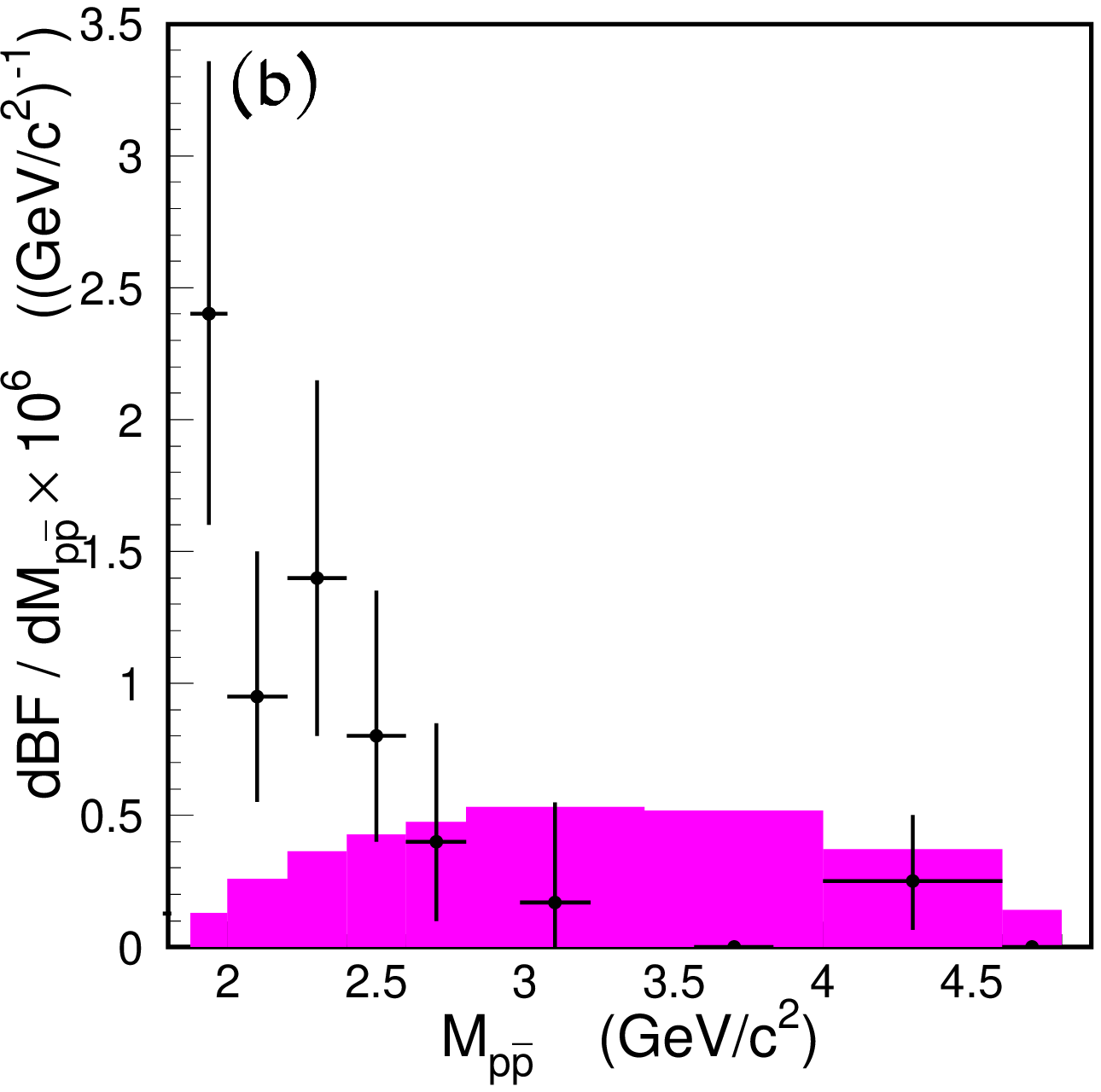,height=1.5in}
\epsfig{figure=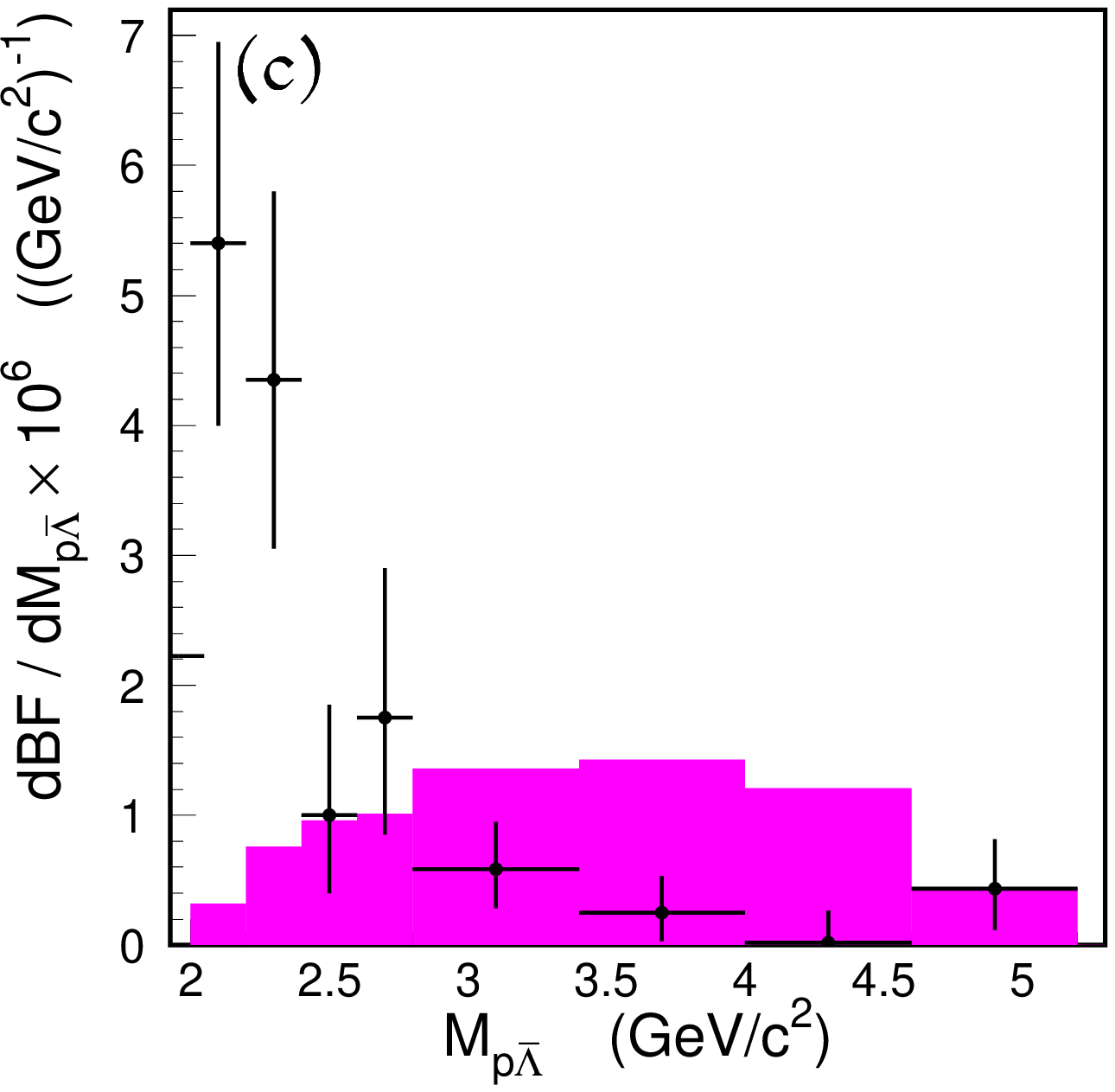,height=1.5in}
\epsfig{figure=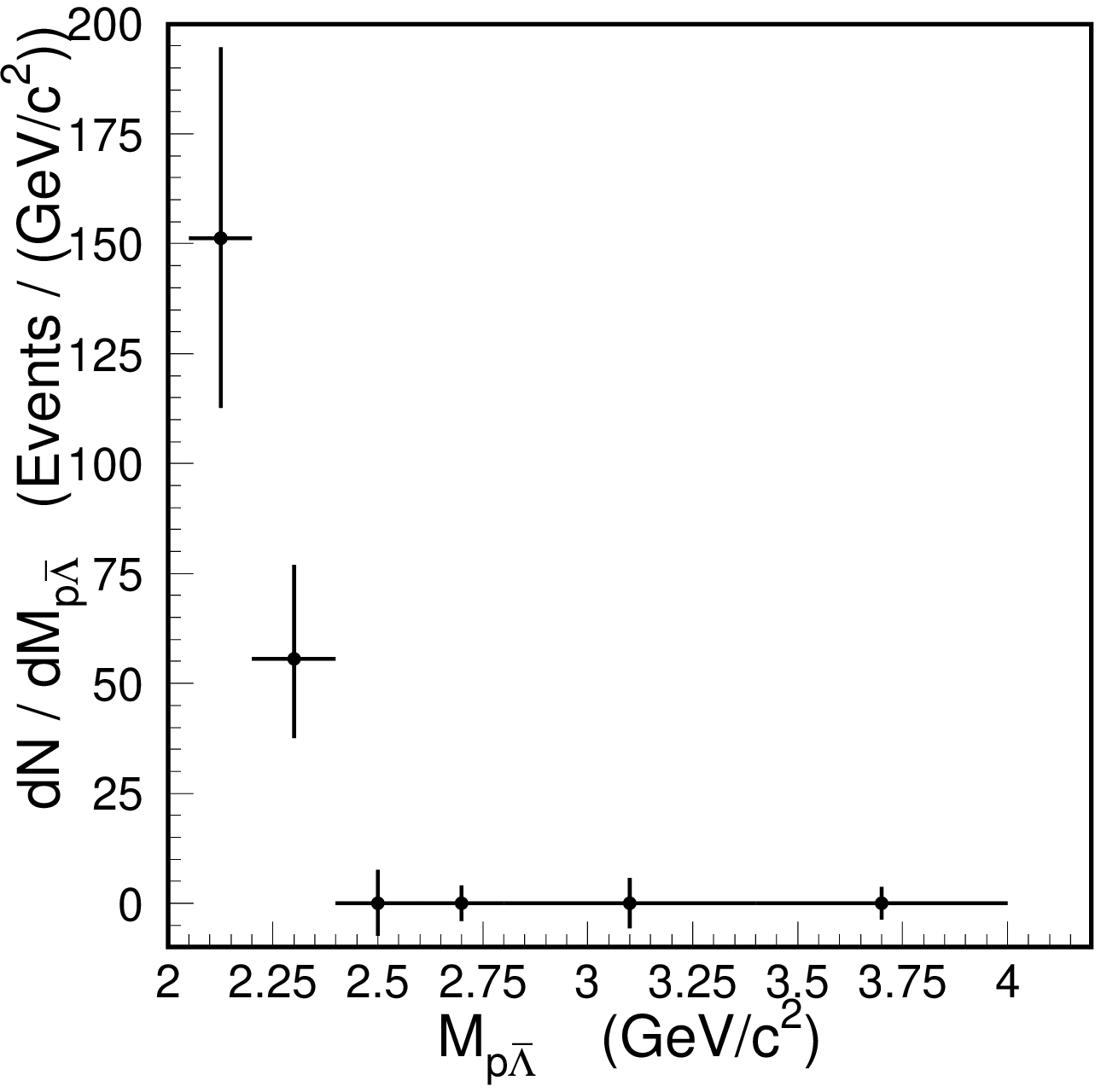,height=1.5in}
\caption{Differential branching fraction for (a) $p\bar{p}K^+$, (b) $p\bar{p}K_S^0$, (c) $p \bar{\Lambda}\pi^-$ and (d) $p\bar{\Lambda}\gamma$
modes as a function of di-baryon pair mass. The shaded distribution shows the 
expectation from a phase-space MC simulation with area scaled to the signal yield. A charmonium veto has been applied in (a) and (b).
\label{fig:dim}}
\end{figure}

Observations of several charmless baryonic $B$ decays have been reported at 
Belle. One common feature of these observations is
the peaking of the di-baryon mass spectra toward threshold. We first measured 
the differential branching fractions for 
(a)$B^+ \to p \bar{p}K^+$, (b)$B^0 \to p \bar{p} K^0_S$, (c)$B^0 \to p \bar{\Lambda}\pi^-$ and (d)$B^+ \to p \bar{\Lambda} \gamma$ modes. There are two kinematic variables in the center
of mass frame which are usually used to extract the $B$ candidates: the beam 
energy constrained mass $M_{\rm bc}$ = $\sqrt{E_{\rm beam}-p^2_B}$, and the energy difference $\Delta E$ = $E_B - E_{\rm beam}$, where $E_{\rm beam}$ is the 
beam energy, and $p_B$ and $E_B$ are the momentum and energy of $B$ candidates.
$B$ signal yield is obtained by a 2D fit to the ($M_{\rm bc}$,$\Delta E$) distribution for each bin of di-baryon invariant mass. The efficiency as a function of
the di-baryon mass is based on the MC simulation. The differential branching fraction
for each bin in the di-baryon mass is obtained from the fitted $B$ yield and the
signal efficiency (Fig. \ref{fig:dim}). The branching fraction is a sum of the
differential branching fractions\cite{bar4,bar5}.

\begin{figure}
\centering
\epsfig{figure=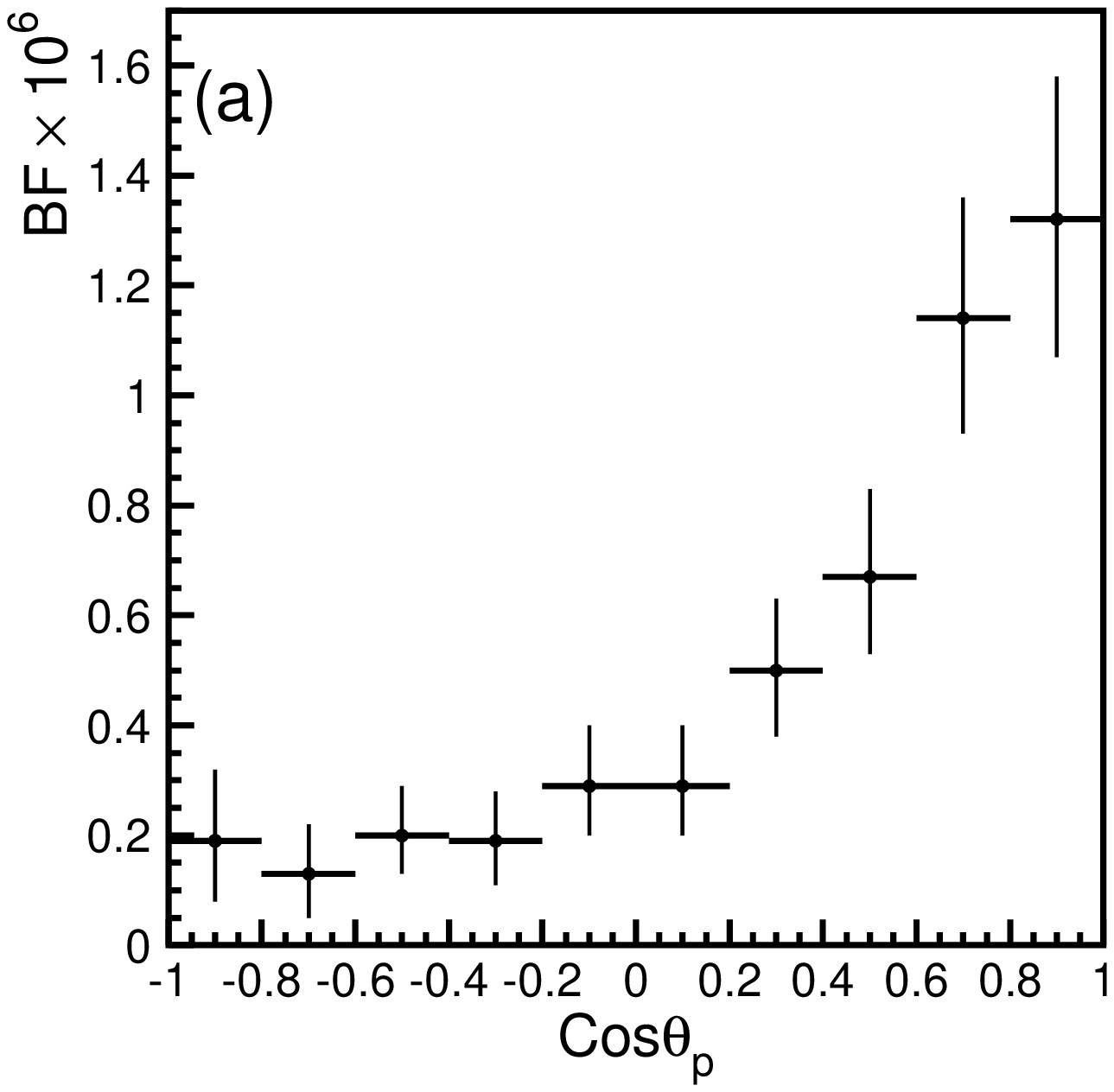,height=1.5in}
\epsfig{figure=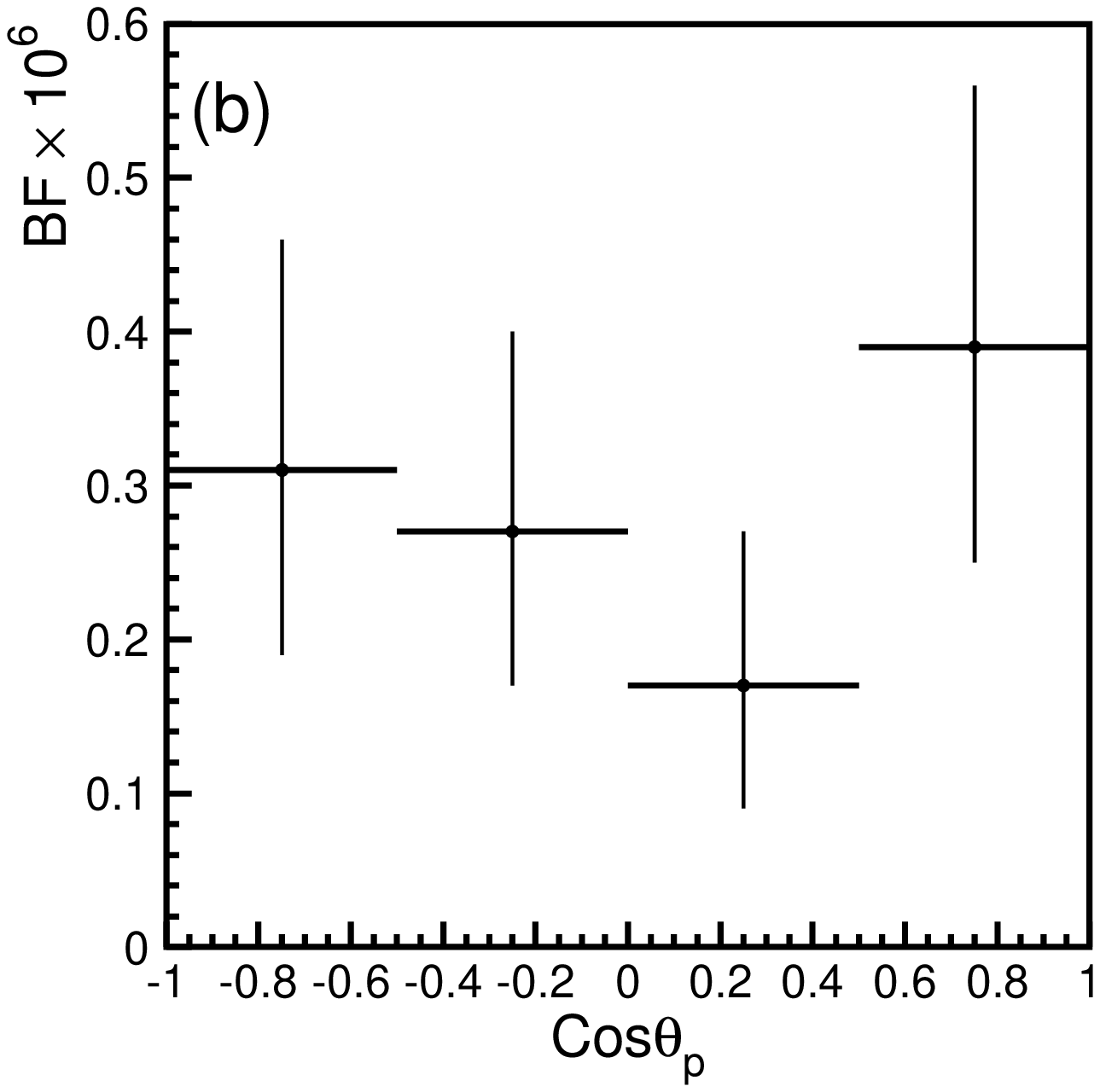,height=1.5in}
\epsfig{figure=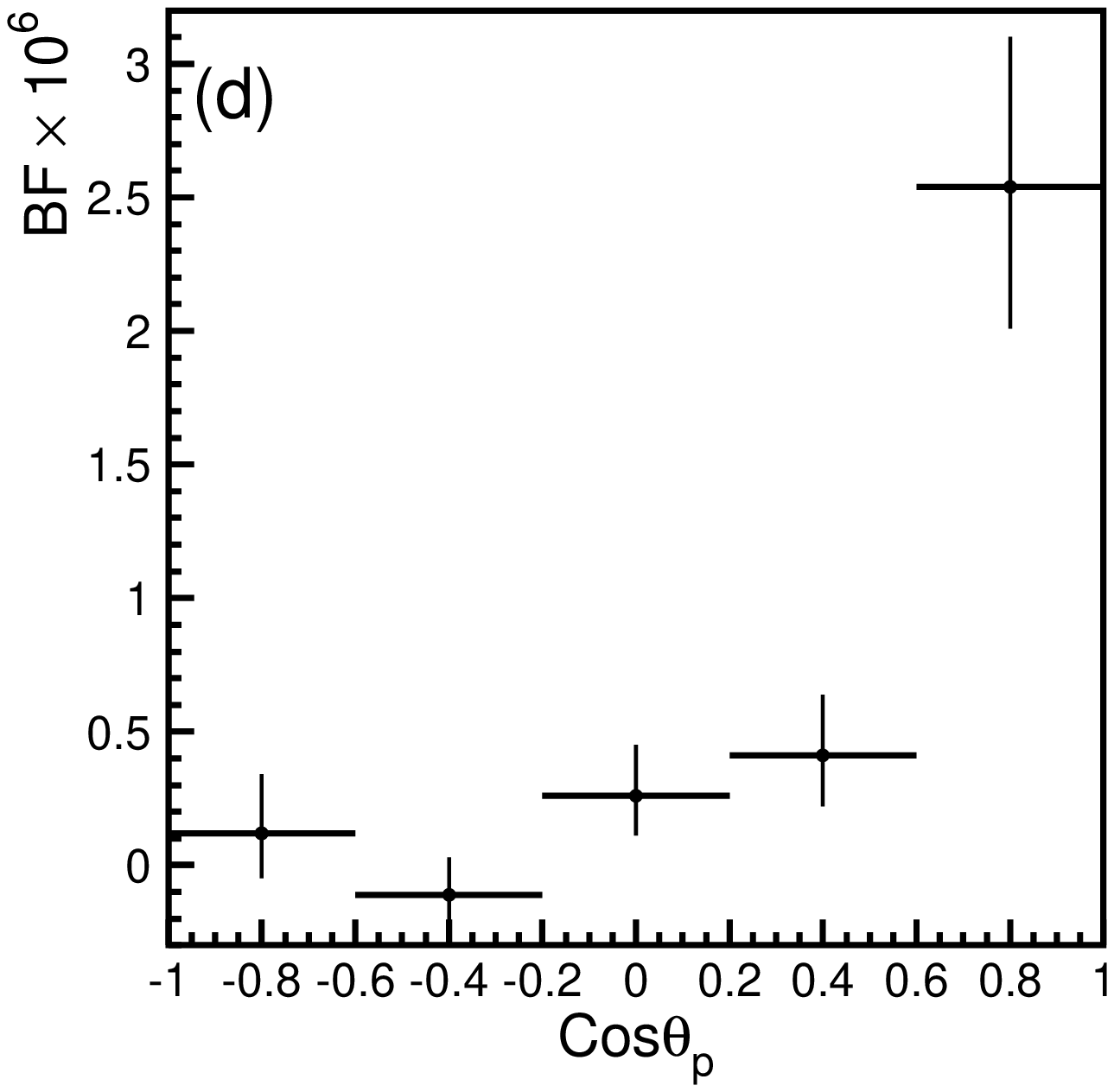,height=1.5in}
\epsfig{figure=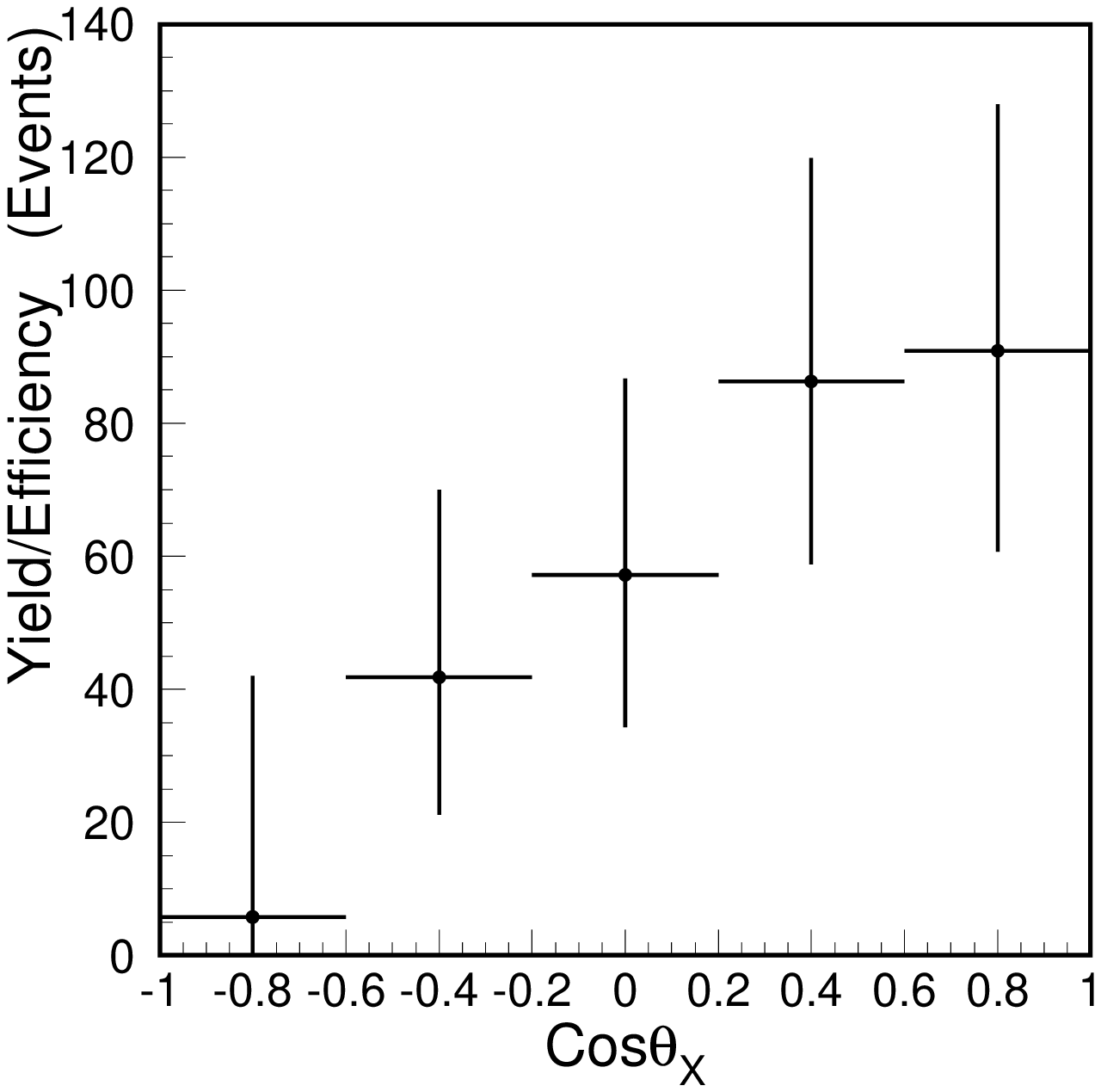,height=1.5in}
\caption{Branching fraction vs. $\cos \theta_p$ in the di-baryon system for (a) $p\bar{p}K^+$, (b) $p\bar{p}K_S^0$, (c) $p \bar{\Lambda}\pi^-$ and (d) $p\bar{\Lambda}\gamma$.
\label{fig:ang}}
\end{figure}

The angular distribution of the proton is studied in the di-baryon system. 
Fig. \ref{fig:ang} shows the angular distributions for the four decay channels. The angle $\theta_p$ is defined for (a) as the angle
between the directions of ${\bar p}$ and  $K^+$ in the $p\bar{p}$ rest frame.
There is a clear forward peak and the angular asymmetry,
defined as
\begin{equation}
A = \frac{N_+-N_-}{N_++N_-}
\end{equation}
where $N_+$ and $N_-$ stand for the efficiency corrected $B$ yield with
$\cos \theta_p>$ 0 and $\cos \theta_p<$0, respectively,
amounts to 0.59$^{+0.08}_{-0.07}$ for the $p \bar{p} K^+$ mode. The asymmetry of the 
distribution indicates that the fragmentation picture is favored. Antiprotons
are emitted along the $K^+$ direction most of the time, which can be explained 
by a parent $\bar{b} \to \bar{s}$ penguin transition followed by 
$\bar{s}u$ fragmentation into the final state. 
The 
$\cos \theta_p$ distribution for (b) is flat, but we have to note that in this 
case the statistics is low and $K_S^0$ carries no flavor information. If in (c) we choose the $\theta_p$ angle of $p$ with respect to the $\pi^-$ in the rest frame of the
$\bar{\Lambda} p$ system, the $\cos \theta_p$ distribution is quite flat. However, the $\cos \theta_p$ shows a forward peak structure if the angle is defined as the $p$ direction relative to the
$\bar{\Lambda}$ direction in the $p\pi^-$ rest frame. It is evident that
the fragmentation interpretation is supported: the proton tends to emerge parallel to the $\bar{\Lambda}$ baryon. The angle $\theta_X$ 
of (d) is measured between the proton direction and the $\gamma$ direction 
in the 
baryon pair rest frame. There is also a clear forward structure in the $\cos \theta_X$ distribution and the angular asymmetry A is 0.36$^{+0.23}_{-0.20}$. This
distribution supports the $b \to s \gamma$ fragmentation picture where the $\Lambda$ tends to emerge opposite to the direction of the photon\cite{bar5}.

\section{Charmed baryonic $B$ decays}
In the $B^{+/0} \to \bar{\Xi}_c^{0/-}\Lambda_c^+$ analysis\cite{xi}, we reconstruct the 
following decay modes: $\Xi_c^0 \to \Xi^- \pi^+$ and $\Lambda K^- \pi^+$, $\Xi_c^+ \to \Xi^- \pi^+ \pi^+$, $\Lambda_c^+ \to p K^- \pi^+$, $\Xi^- \to \Lambda \pi^-$ and $\Lambda \to p \pi^-$. We use a simultaneous two-dimensional binned 
maximum likelihood fit to the $\Delta E$ vs. $M_{\rm bc}$ distributions 
(for the two $\Xi_c^0$ channels) with a common value of ${\cal B}(B^+ \to \bar{\Xi}_c^0 \Lambda_c^+) \times {\cal B}(\bar{\Xi}_c^0 \to \bar{\Xi}^+ \pi^-) \times {\cal B}(\Lambda_c^+ \to p K^- \pi^+)$. For this fit, we constrain the ratio 
${\cal B}(\Xi_c^0 \to \Lambda K^- \pi^+)/{\cal B}(\Xi_c^0 \to \Xi^- \pi^+)$ to 
the recent Belle measurement\cite{les} of $1.07 \pm 0.12 \pm 0.07$. The fit 
gives the product of branching fractions of ${\cal B}(B^+ \to \bar{\Xi}^0_c\Lambda_c^+)\times {\cal B}(\bar{\Xi}_c^0 \to \bar{\Xi}^+\pi^-)$ with the value of 4.8 $^{+1.0}_{-0.9} \pm 1.1 \pm 1.2$ and a 
statistical significance of 8.7 $\sigma$. We also check the 
$B^0 \to \bar{\Xi}_c^-\Lambda_c^+$ mode which is an isospin partner of the 
$B^+ \to \bar{\Xi}^0_c \Lambda_c^+$ mode. The product of branching fraction of 
${\cal B}(B^0 \to \bar{\Xi}^-_c \Lambda^+_c)\times {\cal B}(\bar{\Xi}^-_c \to \bar{\Xi}^+ \pi^- \pi^-)$ is measured to be (9.3$^{+3.7}_{-2.8} \pm 1.9 \pm 2.4$) $\times 10^{-5}$. The uncertainties in the products of branching ratios are 
statistical, systematic and the uncertainty from the $\Lambda_c^+ \to p K^- \pi^+$
branching fraction.

\begin{figure}
\centering
\epsfig{figure=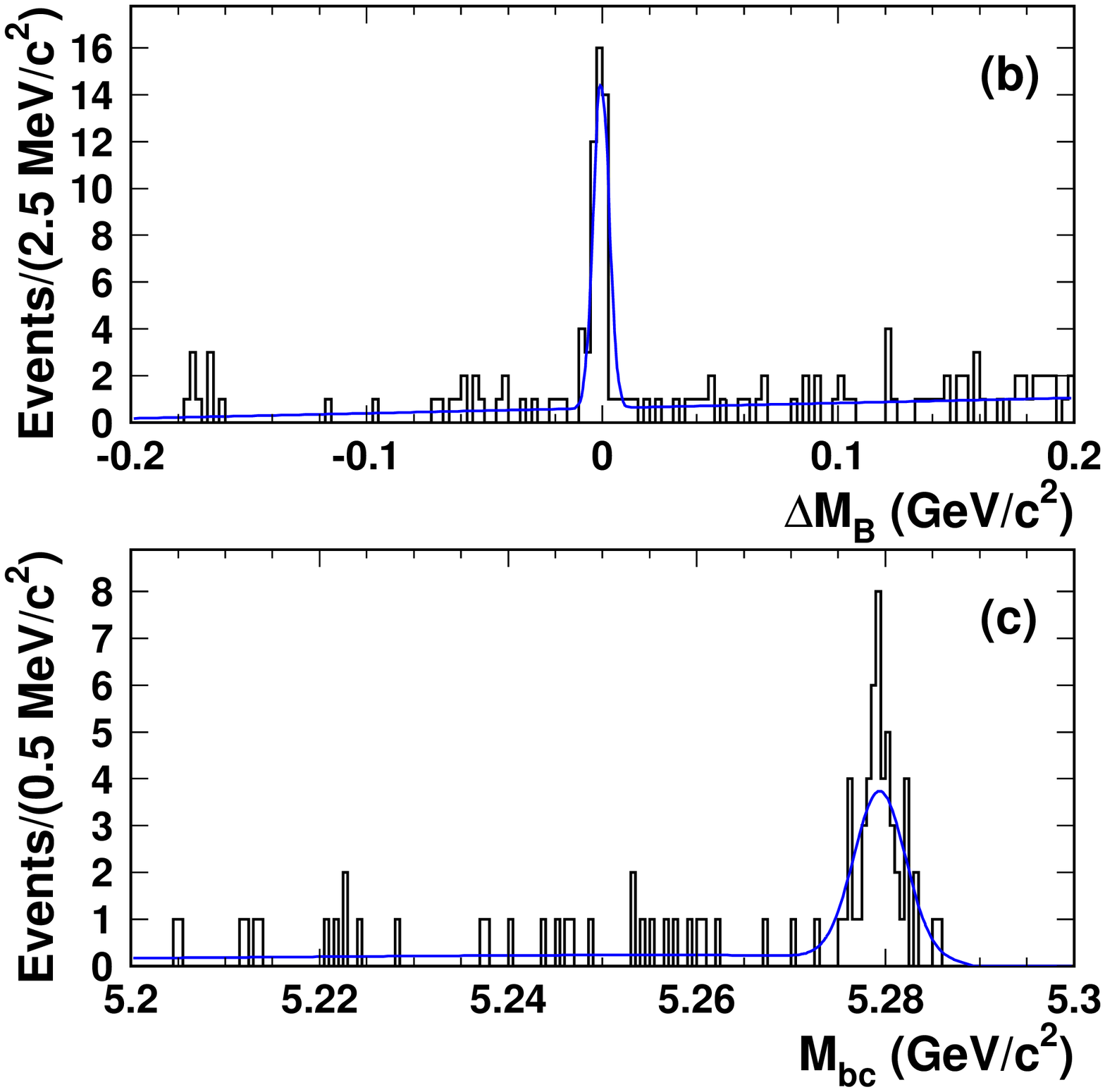,height=1.5in}
\epsfig{figure=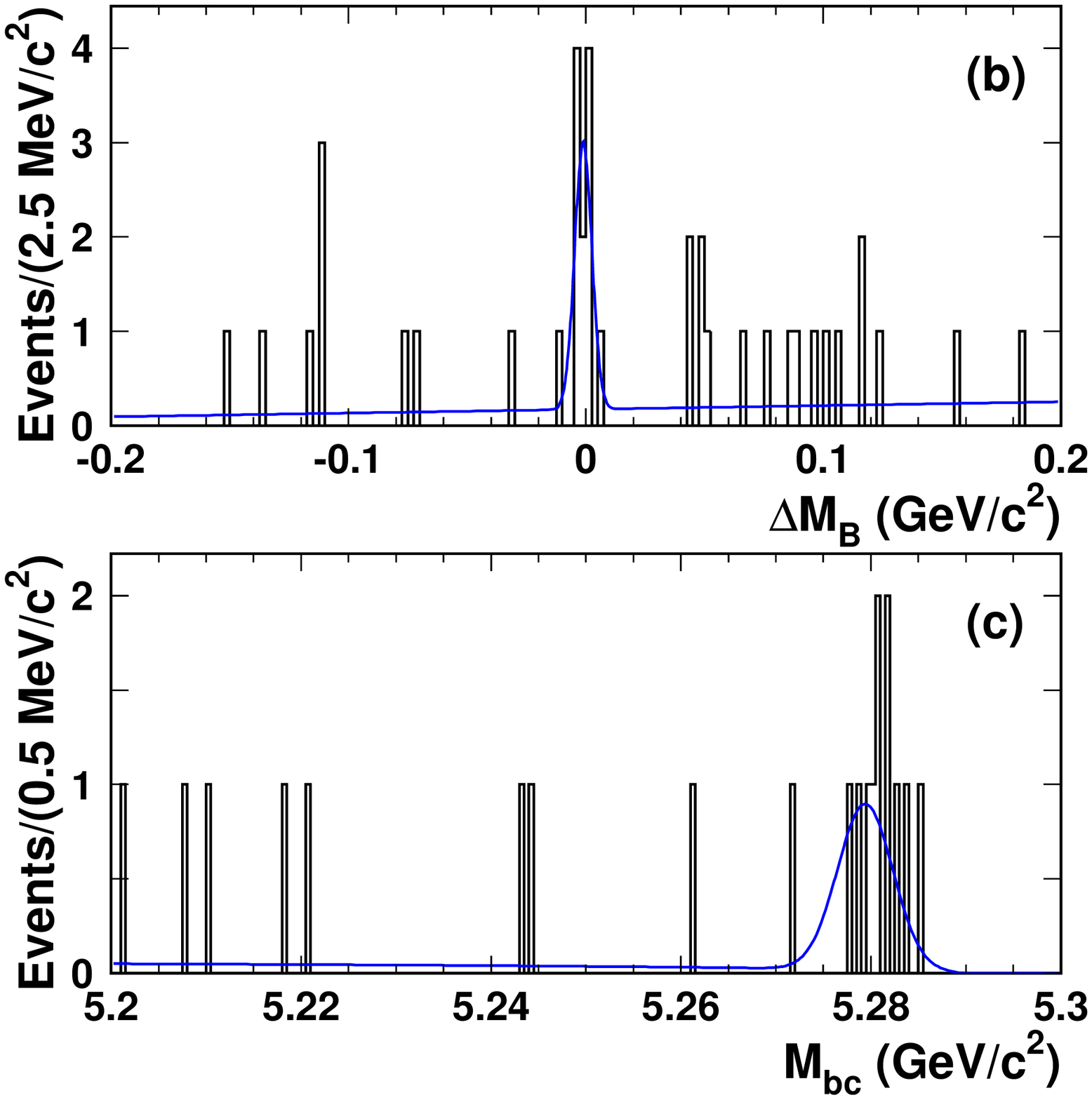,height=1.5in}
\caption{Candidate (a,b) $B^+ \to \Lambda_c^+ \Lambda_c^- K^+$ and (c,d) $B^0 \to \Lambda_c^+ \Lambda_c^- K^0$ decay events: (a,c) $\Delta M_B$ distribution for $M_{\rm bc}>$ 5.27 GeV/c$^2$ and (b,d) $M_{\rm bc}$ distribution for $|\Delta M_B|<$ 0.015 GeV/c$^2$. Curves indicate the fit results.
\label{fig:mb}}
\end{figure}

The $B^+ \to \Lambda_c^+ \Lambda_c^- K^+$ and $B^0 \to \Lambda_c^+ \Lambda_c^- K^0$ decays are three-body decays that proceed via a $b \to c\bar{c}s$ 
transition. We detect the $\Lambda_c^+$ via the $\Lambda_c^+ \to p K^- \pi^+$,
$p \bar{K}^0$ and $\Lambda\pi^+$ decay channels. When a $\Lambda_c^+$ and 
$\Lambda_c^-$ are combined as $B$ decay daughters, at least one of $\Lambda_c^{\pm}$ is required to have been reconstructed via the $p K^{\mp} \pi^{\pm}$ decay 
process. Here, the parameter mass difference $\Delta M_B$ is used instead of the
energy difference $\Delta E$, since $\Delta E$ shows a correlation with $M_{\rm bc}$. The mass difference is defined as $\Delta M_B \equiv M(B) - m_B$, where 
$M(B)$ is the reconstructed mass of the $B$ candidate and $m_B$ is the world
average $B$ meson mass. Fig. \ref{fig:mb} shows $\Delta M_B$ and $M_{\rm bc}$
projections for $B^+ \to \Lambda_c^+ \Lambda_c^- K^+$ and $B^0 \to \Lambda_c^+ \Lambda_c^- K^0$ decays. A two-dimensional binned maximum likelihood fit is
performed to determine the signal yield. From the fit we obtain signal yields of
48.5$^{+7.5}_{-6.8}$ and 10.5$^{+3.8}_{-3.1}$ events with statistical 
significances of 15.4 $\sigma$ and 6.6 $\sigma$, for $B^+ \to \Lambda_c^+ \Lambda_c^- K^+$ and $B^0 \to \Lambda_c^+ \Lambda_c^- K^0$, respectively. Hence, we 
obtain the branching fractions of\cite{gab} 
\bea
{\cal B}(B^+ \to \Lambda_c^+ \Lambda_c^- K^+) & = & (6.5^{+1.0}_{-0.9} \pm 1.1 \pm 3.4) \times 10^{-4} \nonumber \\
{\cal B}(B^0 \to \Lambda_c^+ \Lambda_c^- K^0) & = & (7.9^{+2.9}_{-2.3} \pm 1.2 \pm 4.1) \times 10^{-4} \nonumber
\label{eq:sp}
\eea
where the first and the second errors are statistical and systematic, respectively. The last error is due to the 52 \% uncertainty in the absolute branching fraction, ${\cal B}(\Lambda_c^+ \to p K^-\pi^+)$.

We study two-body baryonic decays of charmonia in the $B$ meson decays,
$B^+ \to p\bar{p}K^+$ and $B^+ \to \Lambda \bar{\Lambda}K^+$. The $B$ signal yields are obtained from 10 MeV/c$^2$ wide $M_{p\bar{p}}$($M_{\Lambda\bar{\Lambda}}$) mass bins from the kinematic threshold to 4.5
GeV/c$^2$. 
The results of $\eta_c$ are the mass of M$_{\eta_c}$ = 2.971 $\pm$ 0.003$^{+0.002}_{-0.001}$ GeV/c$^2$(2.974$\pm$0.007$^{+0.002}_{-0.001}$ GeV/c$^2$) and the
width of $\Gamma(\eta_c)$ = 48$^{+8}_{-7}$ $\pm$ 5 MeV/c$^2$(40$\pm$ 19 $\pm$ 5
MeV/c$^2$) from $\eta_c \to p \bar{p}$($\eta_c \to \Lambda \bar{\Lambda}$) mode.
We define the $\eta_c$ signal region as 2.940 GeV/c$^2$ $<$ $M_{\Lambda \bar{\Lambda}}$ $<$ 3.020 GeV/c$^2$. The fitted $B$ signal yield, efficiency and
branching fraction are shown in Table \ref{tab:jpsi}. In this study the decay 
$\eta_c \to \Lambda \bar{\Lambda}$ has been observed for the first time, with ${\cal B}(\eta_c \to \Lambda \bar{\Lambda}) = (0.87^{+0.24+0.09}_{-0.21-0.14}\pm 0.27)\times 10^{-3}$. The observed ${\cal B}(\eta_c \to \Lambda \bar{\Lambda})/{\cal B}(\eta_c \to p \bar{p})$ is $0.67^{+0.19}_{-0.16} \pm 0.12$ which is consistent with theoretical expectation\cite{ans}. We define the $J/\psi$ signal region as 3.075 GeV/c$^2$ $<$ $M_{p\bar{p}}$($M_{\Lambda \bar{\Lambda}}$) $<$ 3.117 GeV/c$^2$ and use events
in this signal region to study the proton angular distribution in the helicity
frame of the $J/\psi$. The helicity angle $\theta_X$ is defined as the angle
between the proton flight direction and the direction opposite to the flight of
the kaon in the $J/\psi$ rest frame. The angular distribution of the kaon 
direction in the $J/\psi$ rest frame
is parameterized as $P(\alpha_B,\cos \theta_X) = (1+\alpha_B \cos^2 \theta_X)/(2+2\alpha_B/3)$ with $\alpha_B =(-2\alpha)/(\alpha+1)$. We determine\cite{mur} $\alpha_B$ to be -0.60 $\pm$ 0.13 $\pm$ 0.14 ($p\bar{p}$) and -0.44 $\pm$ 0.51 $\pm$ 0.31 ($\Lambda \bar{\Lambda}$.
\begin{table}[t]
\caption{Measured Branching Fractions ${\cal B}(J/\psi,\eta_c \to p\bar{p},\Lambda \bar{\Lambda})$.\label{tab:jpsi}}
\vspace{0.4cm}
\begin{center}
\begin{tabular}{|c|c|c|c|}
\hline
Modes & Yield & Efficiency(\%) & ${\cal B}(J/\psi,\eta_c \to p\bar{p},\Lambda \bar{\Lambda}) \times 10^{-3}$ \\
\hline
$B^+ \to \eta_c K^+,\eta_c \to p\bar{p}$ & 195.1$^{+15.7}_{-14.7}$ & 35.8$\pm$ 0.3 & 1.58 $\pm$ 0.12 $^{+0.18}_{-0.22}$ $\pm$ 0.47 \\
$B^+ \to \eta_c K^+,\eta_c \to \Lambda \bar{\Lambda}$ & 19.5$^{+5.2}_{-4.5}$ & 5.3 $\pm$ 0.1 & 0.87$^{+0.24+0.09}_{-0.21-0.14} \pm 0.27$ \\
$B^+ \to J/\psi K^+, J/\psi \to p\bar{p}$ & 317.2$^{+19.0}_{-18.0}$ & 37.3 $\pm$0.4 & 2.21 $\pm$ 0.13 $\pm$ 0.31 $\pm$ 0.10 \\
$B^+ \to J/\psi K^+, J/\psi \to \Lambda \bar{\Lambda}$ & 45.9$^{+7.7}_{-6.7}$ & 5.9$\pm$ 0.3 & 2.00$^{+0.34}_{-0.29}\pm$ 0.34 $\pm$ 0.08 \\
\hline
\end{tabular}
\end{center}
\end{table}

\end{document}